\documentclass[aps,prl,twocolumn,showpacs,superscriptaddress,groupedaddress]{revtex4}  
\usepackage{graphicx}  
\usepackage{dcolumn}   
\usepackage{bm}        
\usepackage{amssymb}   
\usepackage{amsmath}   

\usepackage{verbatim}  

\begin{document}
\title{Quantum ammeter}

\author{Edvin G. Idrisov}
\affiliation{Physics and Materials Science Research Unit, University of Luxembourg, L-1511 Luxembourg, Luxembourg}

\author{Ivan P. Levkivskyi}
\affiliation{ Dropbox Ireland, One Park Place, Hatch Street Upper, Dublin, Ireland}

\author{Eugene V. Sukhorukov}
\affiliation{D\'epartement de Physique Th\'eorique, Universit\'e de Gen\`eve, CH-1211 Gen\`eve 4, Switzerland}
\date{\today}




\begin{abstract}
We present the theoretical model of the ``quantum ammeter'', a device that is able to measure the full counting statistics of an electron current at quantum time scales. It consists of an Ohmic contact, perfectly coupled to chiral quantum Hall channels, and of a quantum dot attached to one of the outgoing channels. At energies small compared to its charging energy, the Ohmic contact fractionalizes each incoming electron and redistributes it between outgoing channels. By monitoring the resonant tunneling current through the quantum dot, one gets an access to the moment generator of the current in one of the incoming channels at time scales comparable to its correlation time.        
\end{abstract}

\pacs{73.21.-b, 73.43.-f, 73.43.Lp, 73.63.-b, 85.35.-p}
\maketitle


Since their discovery, the integer \cite{Klitzing} and fractional \cite{Tsui} quantum Hall (QH) effect,  admittedly being a very complex phenomena \cite{Ezawa}, for many years remained predominantly a playground for various theoretical models. For instance, although Lauglin quasi-particles with fractional charge \cite{Laughlin} were successfully observed in experiments based on shot noise measurements \cite{Reznikov,Saminadayar}, their coherence and anyonic exchange statistics are still illusive. Only recently, we have witnessed a number of thorough experimental studies exploring different aspects of the integer QH effect physics, such as the quantum coherence of the edge states \cite{Ji}, the energy relaxation and heat transport at the QH edge \cite{PPPFujisawa}, 
and many others.

A further progress in experimental techniques at nanoscale provided one with the remarkable control of the electronic quantum states in QH systems, thereby giving birth to the new field in the QH effect physics, dubbed the {quantum electron optics} \cite{Degiovanni}. In this field, experimentalists combine different nanoscale systems, such as quantum point contacts (QPC), quantum dots (QD), QH edge channels, to study specific quantum phenomena.
Between these basic elements of the electron quantum optics, an Ohmic contact, a piece of metal perfectly coupled to QH edge channels, was always considered merely a reservoir of equilibrium electrons \cite{Buttiker}, i.e., an analog of a black body in quantum optics. However, in contrast to photons electrons strongly interact. As a consequence, if the capacitance $C$ of an Ohmic contact is relatively small, so that its charging energy $E_{c}=e^2/2C$ is not negligible, new interaction effects arise.
The notable examples are the suppression of charge quantization caused by quantum fluctuations \cite{Pierre1,Idrisov1}, heat Coulomb blockade (CB) effect \cite{Slobodeniuk,Pierre2}, interaction induced recovery of the phase coherence \cite{Clerk, Idrisov2,Pierre4}, charge Kondo effect \cite{Matveev, Pierre3}, quantization of the anyonic heat flow \cite{Heiblum1}, and the observation of the half-integer thermal Hall conductance \cite{Heiblum2}.

\begin{figure}
\includegraphics[scale=.6]{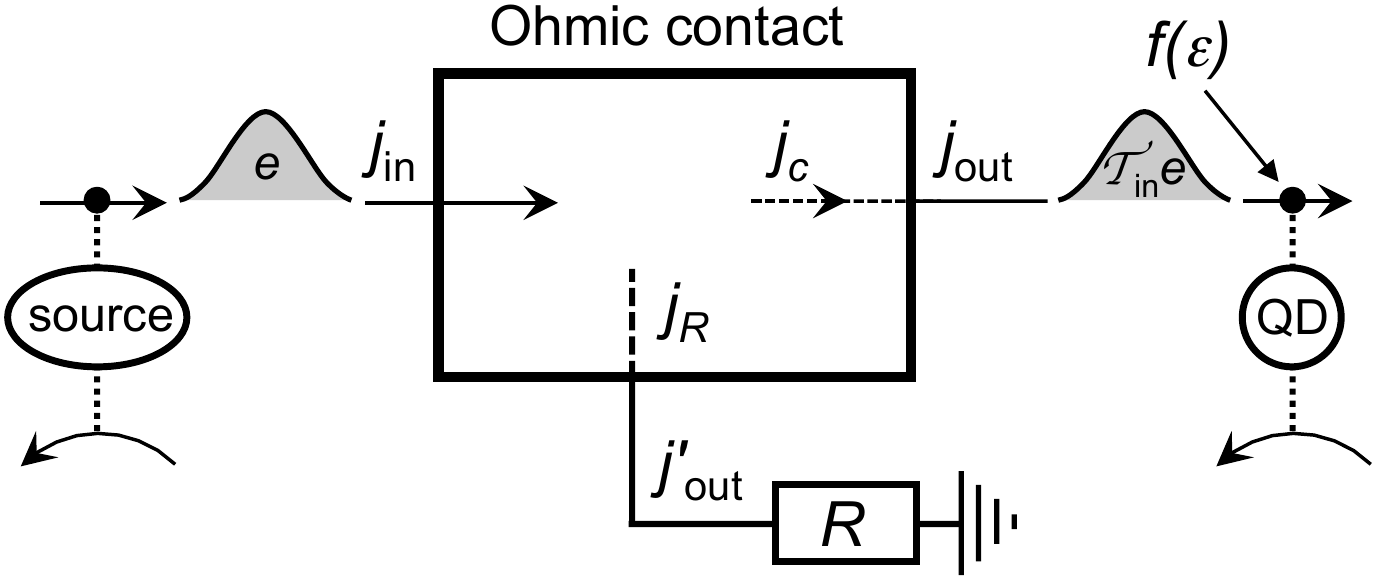}
\caption{\label{fig:one} The quantum ammeter, a system that can be used to measure an electron current fluctuations (injected from the source) at quantum time scales, is schematically shown. It consists of an Ohmic contact (a small piece of metal containing the charge $Q$) perfectly coupled to QH edge states at integer filling factor and loaded by the resistance $R$, and of a QD attached to the outgoing channel. The Ohmic contact splits every incoming electron and transmits ${\cal T}_{{\rm in}}=R/(R+R_q)$ part of it to the outgoing channel, where $R_q$ is the quantum of resistance. The electron distribution function in the outgoing channel $f(\varepsilon)$ is measured by monitoring the resonant tunneling current through a QD, as in Ref.\ [\onlinecite{PPPFujisawa}]. The function $f(\varepsilon)$ contains the contribution from the moment generation function of the current  $j_{{\rm in}}$ injected in the incoming channel.  The constant $\lambda_{{\rm in}}=2\pi {\cal T}_{{\rm in}}$ plays the role of a counting variable in the moment generator, which can be controlled by the the resistance $R$. We assume, that the load resistor does not create partition noise, which is the case where, e.g., it fully transmits $N$ non-chiral modes, so that $R=R_q/N$. Other notations are explained in the text of the paper.}
\end{figure}

In this Letter, we consider an Ohmic contact as an ideal electron fractionalizer. By this we mean that at energies smaller than its charging energy  $E_c$, the Ohmic contact simply splits every electron injected in the incoming channel so that only the fraction ${\cal T}_{{\rm in}}\leq 1$ of the incoming electron charge $e$ is transmitted to one of the outgoing channel, as shown in Fig.\ \ref{fig:one}. This process, caused by the strong Coulomb interaction at the Ohmic contact, is fully deterministic and has to be distinguished from electron beam splitting, where the outgoing state is a superposition of the full (not fractionalized) electron in the outgoing channels.

We furthermore propose to use this system in conjunction with a QD as a  ``quantum ammeter'', i.e., a device that is able to detect the full counting  statistics (FCS) of the fluctuations of the injected current at quantum time scales \cite{Levitov}. Namely, by measuring the resonant tunneling current through a QD, on can record the electron distribution function $f(\varepsilon)$ in the outgoing channel (as experimentally demonstrated in Ref.\ [\onlinecite{PPPFujisawa}]), which is directly related to the electron correlation function in the channel, see Eq.\ (\ref{distrib}). Every time an injected electron crosses the tunneling point, it adds the Friedel's phase shift $\lambda_{{\rm in}}=2\pi {\cal T}_{{\rm in}}$ to the electron correlation function, which after time $t$ acquires the overall contribution in the form of the moment generating function of the charge (\ref{result}) injected to the incoming channel \cite{footnote0}.  The parameter $\lambda_{{\rm in}}$ plays the role of the counting variable, which can be controlled by a load resistor $R$, ideally without adding extra partition noise. In the following, we present the theory of the quantum ammeter and discuss its limitations. Throughout the paper, we set $|e|=\hbar=k_B=1$, which also implies that the resistance quantum $R_q=2\pi\hbar/e^2=2\pi$.

\textit{Model of the quantum ammeter}. We follow Refs.\ \cite{Matveev, Slobodeniuk, Idrisov1,Idrisov2} and use the effective theory \cite{Wen} and the bosonization technique \cite{Giamarchi} to describe QH edge states perfectly coupled to the Ohmic contact. Accordingly, the collective fluctuations of the charge densities
$\rho_{\alpha}(x,t)$ and currents $j_{\alpha}(x,t)$ in the QH edge channels are expressed in terms of the bosonic fields $\phi_{\alpha}(x,t)$,
\begin{equation}
\rho_{\alpha}=(1/2\pi)\partial_x \phi_{\alpha},\quad
j_{\alpha}=-(1/2\pi)\partial_t \phi_{\alpha},
\label{fields}
\end{equation} 
where the index $\alpha$ enumerates channels, entering and leaving the Ohmic contact (see Fig.\ \ref{fig:one}).  The electrons in the channels are represented by the vertex operators, $\psi_\alpha(x,t)\propto e^{i\phi_\alpha(x,t)}$, so that the electron distribution function in the outgoing channel reads:
\begin{equation}
f(\varepsilon) = \int\limits_{-\infty}^\infty \frac{dt}{2\pi } e^{-i\varepsilon t} K(t), \quad K(t) \propto \langle e^{-i\phi_{{\rm out}}(t)}e^{i\phi_{{\rm out}}(0)}\rangle.
\label{distrib}
\end{equation}
Here, the prefactor in the expression for $K(t)$ can be fixed in the end of calculations by comparing to the equilibrium correlation function for free electrons.

Next, we apply the Langevin equation method \cite{Slobodeniuk,Sukhorukov1} to evaluate electron correlation function $K(t)$. This method has been successfully used \cite{Idrisov1,Pierre1} to describe experiments on the decay of CB oscillations \cite{Pierre1,Pierre2,Pierre3}. According to this method, an Ohmic contact serves as a resorvoir of neutral excitations, which are accounted by introducing Langevin currents $j_{c}$ and $j_{R}$ in the equation of motion for the charge $Q$ of the contact (see Fig.\ref{fig:one}). The details of the evaluations are presented in the supplementary material, while the results are rather intuitive. At frequencies small compared to the charging energy, $\omega\ll E_{c}$, the outgoing current is simply expressed in terms of the incoming current  $j_{{\rm in}}$, and the Langevin sources, $j_{c}$ and $j_{R}$:
\begin{equation}
\label{Outgoing currents}
 j_{{\rm out}}(\omega)=\sum_{\alpha} {\cal T}_\alpha j_\alpha(\omega),\quad
  j'_{\rm out}(\omega)=\sum_{\alpha} {\cal T}'_\alpha j_\alpha(\omega),
\end{equation}
where $\alpha=({\rm in},c,R)$, and the frequency-independent transmission coefficients take the following form
\begin{subequations}
\label{coefficients}
\begin{align}
\label{coefficients1}
& -{\cal T}_{R}={\cal T}_{{\rm in}}=1-{\cal T}_{c}=R/(R_q+R),\\
& -{\cal T}'_{c}={\cal T}'_{{\rm in}}=1-{\cal T}'_{R}=R_q/(R_q+R).
\label{coefficients2}
\end{align}
\end{subequations}
By solving Eqs.\ (\ref{fields}-\ref{coefficients}), one  arrives at the following  expression for the electron correlation function 
\begin{equation}
K(t)\propto \prod_{\alpha= {\rm in},c,R}\langle e^{i\lambda_\alpha Q_\alpha(t)}e^{-i\lambda_\alpha Q_\alpha(0)}\rangle,
\label{K}
\end{equation}
where $Q_\alpha (t)=\int^{t}_{-\infty} j_\alpha (t^{\prime})dt^{\prime}$ is the total charge in the channel $\alpha$, and  $\lambda_{\alpha}=2\pi {\cal T}_{\alpha}$. 

At zero bath temperature the Langevin sources $j_\alpha$ acquire  correlation functions in the form
\begin{subequations}
\label{LS}
\begin{align}
\label{LS1}
& \langle j_\alpha(\omega) j_\alpha(\omega')\rangle=2\pi S_\alpha(\omega)\delta(\omega+\omega'), \quad \alpha= c,R,\\
& S_{c}(\omega)=\frac{\omega/R_q}{1-e^{-\omega/T_{c}}},\quad S_{R}(\omega) =\frac{\omega/R}{1-e^{-\omega/T_{c}}}+\frac{\omega\theta(\omega)}{R}, 
\label{LS2}
\end{align}
\end{subequations}
where $T_{c}$ is the temperature of the Ohmic contact. The current $j_{c}$ is chiral, while $j_{R}$ is not, therefore the second term in the expression for $S_{R}$ is the contribution of the incoming states. These relations follow from the fluctuation-dissipation theorem for currents.
Assuming for the moment a cold Ohmic contact, $T_{c}=0$ (the heating effects are studied below),
and using Eqs.\ (\ref{fields}-\ref{K}), we arrive at the following expression for the distribution function 
\begin{subequations}
\label{result}
\begin{align}
\label{distrib2}
&-\frac{d f(\varepsilon)}{d\varepsilon} = \int _{-\infty}^\infty\frac{dt}{2\pi } e^{-i\varepsilon t} {\cal G}(\lambda_{{\rm in}},t), \\
&{\cal G}(\lambda_{\rm{in}},t)= (t+i0)^{{\cal T}^2_{{\rm in}}}\langle e^{i\lambda_{\rm in} Q_{\rm in}(t)}e^{-i\lambda_{\rm in} Q_{\rm in}(0)}\rangle,
\label{FCS}
\end{align}
\end{subequations}
where the correlation function ${\cal G}(\lambda_{\rm in},t)$, normalized to its ground-state part $1/(t+i0)^{{\cal T}^2_{\rm in}}$,
can be considered the quantum FCS generator \cite{Levitov} of the injected non-equilibrium current $j_{\rm in}$, with $\lambda_{\rm in}=2\pi {\cal T}_{\rm in}$ playing the role of the counting variable \cite{footnote0}. In particular, if no current is injected in the ammeter, then ${\cal G}=1$ and 
$df(\varepsilon)/d\varepsilon=-\delta(\varepsilon)$, as expected.

\textit{Applications.} In what follows, we  consider examples of various systems injecting currents with different statistics into the incoming channel, as shown in Fig.\ (\ref{fig:one}). We start with the simplest example of the tunneling current through a barrier with the transparency $D\ll 1$. At long times, $t\Delta\mu \gg 1$, where $\Delta\mu$ is the applied potential difference, the fluctuations of charge can be considered classical, and their statistics is known to be Poissonian \cite{Levitov}: $\log({\cal G})=(t\Delta\mu/2\pi) D(e^{i\lambda_{\rm in}}-1)$ for $t>0$. Taking into account the fact, that all the odd (even) cumulants are odd (even) functions of time, one can write   
\begin{subequations}
\begin{align}
\label{Lorentzian}
&-\frac{df(\varepsilon)}{d\varepsilon}=\frac{1}{\pi}\frac{\gamma}{(\varepsilon-\varepsilon_0)^2+\gamma^2},\\
&\varepsilon_0 =\frac{\Delta\mu D}{2\pi} \sin(\lambda_{\rm in}),\quad \gamma=\frac{\Delta\mu D}{2\pi} [1-\cos(\lambda_{\rm in})].
\label{parameters}
\end{align}
\end{subequations}
This result is justified in the limit $D\ll 1$, since the integral in Eq.\ ({\ref{distrib}) comes from long (Markovian) time scales. For example, if the load resistor transmits only one channel, $R=R_q$, one has $\lambda_{\rm in}=\pi$, and the broadening acquires the maximum value of $\gamma=\Delta\mu D/\pi$, while the energy shift vanishes, $\varepsilon_0=0$. Let us compare this result to the Gaussian noise case, where the functions in Eq.\ (\ref{parameters}) have to be expanded to second order in $\lambda_{\rm in}$. Then,  the energy shift $\varepsilon_0=\Delta\mu D/2$ is simply induced by the average bias in the outgoing QH edge channel, and the broadening $\gamma=\pi\Delta\mu D/4$ is larger compared to the one for the non-Gaussian noise.  Finally, we note, that the contributions to ${\cal G}$ from short time scales $t\sim1/\Delta\mu$ lead to the asymmetry of the tails of the peak in $-df(\varepsilon)/d\varepsilon$, studied in Refs.\ \cite{Chernii,Borin}.

Next, we consider the classical noise induced by sequential tunneling through a resonant level at zero temperature and with the in and out rate $\Gamma_1$ and $\Gamma_2$, respectively. The FCS generator of this process can be obtained by solving the generalized master equation \cite{deJong} with the result ${\cal G}=\sum_{m=1,2}{\cal G}_m e^{\Lambda_m t}$ for $t>0$, where $\Lambda_m(\lambda_{\rm in})$ are the two eigenvalues of the  transition matrix, and ${\cal G}_m (\lambda_{\rm in})$ are the corresponding weights (see the supplementary material). In the long time limit, one of the exponential function dominates, and the noise becomes Markovian with the known generator \cite{deJong}. 
In the symmetric case, $\Gamma_1=\Gamma_2=\Gamma$, the weights are real functions, ${\cal G}_m=[1\pm\cos(\lambda_{\rm in}/2)]/2$, while eigenvalues take the simple form: $\Lambda_m=-\Gamma\pm\Gamma e^{i\lambda_{\rm in}/2}$. Therefore, the function $-df(\varepsilon)/d\varepsilon$ acquires the double-peak structure,
\begin{equation}
\label{Lorentzian2}
-\frac{df(\varepsilon)}{d\varepsilon}=\frac{1}{\pi}\sum_{m=1,2}\frac{{\cal G}_m\gamma_m}{(\varepsilon-\varepsilon_m)^2+\gamma_m^2},\\
\end{equation}
where  $\varepsilon_m=\pm\Gamma\sin(\lambda_{\rm in}/2)$ and $\gamma_m=\Gamma[1\mp\cos(\lambda_{\rm in}/2)]$. In particular, for $R=R_q$ one has $\lambda_{\rm in}=\pi$, and thus the function $-df/d\varepsilon$ shows two equal peaks with ${\cal G}_m=1/2$, $\gamma_m=\Gamma$, and $\varepsilon_m=\pm \Gamma$.

Finally, we consider the noise of a QPC at intermediate transparencies $D$ as an example of the quantum process. In this case, one may observe the noise induced phase transition, studied in Ref.\ \cite{Levkivskyi}, which originates from the singularity in the Markovian cumulant generator at $D=1/2$ and $\lambda_{\rm in}=\pi$, i.e., $R=R_q$. Here, we concentrate on the Gaussian noise case (for the details of calculations, see the supplementary material) and present the full time dependence of the cumulant generator $\log {\cal G}=it\Delta\mu D{\cal T}_{\rm in}-2 D(1-D){\cal T}^2_{\rm in}F(t\Delta\mu)$, where $F(t\Delta\mu)=\int_0^1dx(1-x)x^{-2}[1-\cos(t\Delta\mu x)]$. We evaluate the time integral in Eq.\ (\ref{distrib}) numerically and show the result in Fig.\ \ref{fig:two} for the transparency of the QPC $D=1/2$ and for the load resistance $R=R_q/2$ (blue line).

\begin{figure}
\includegraphics[scale=.6]{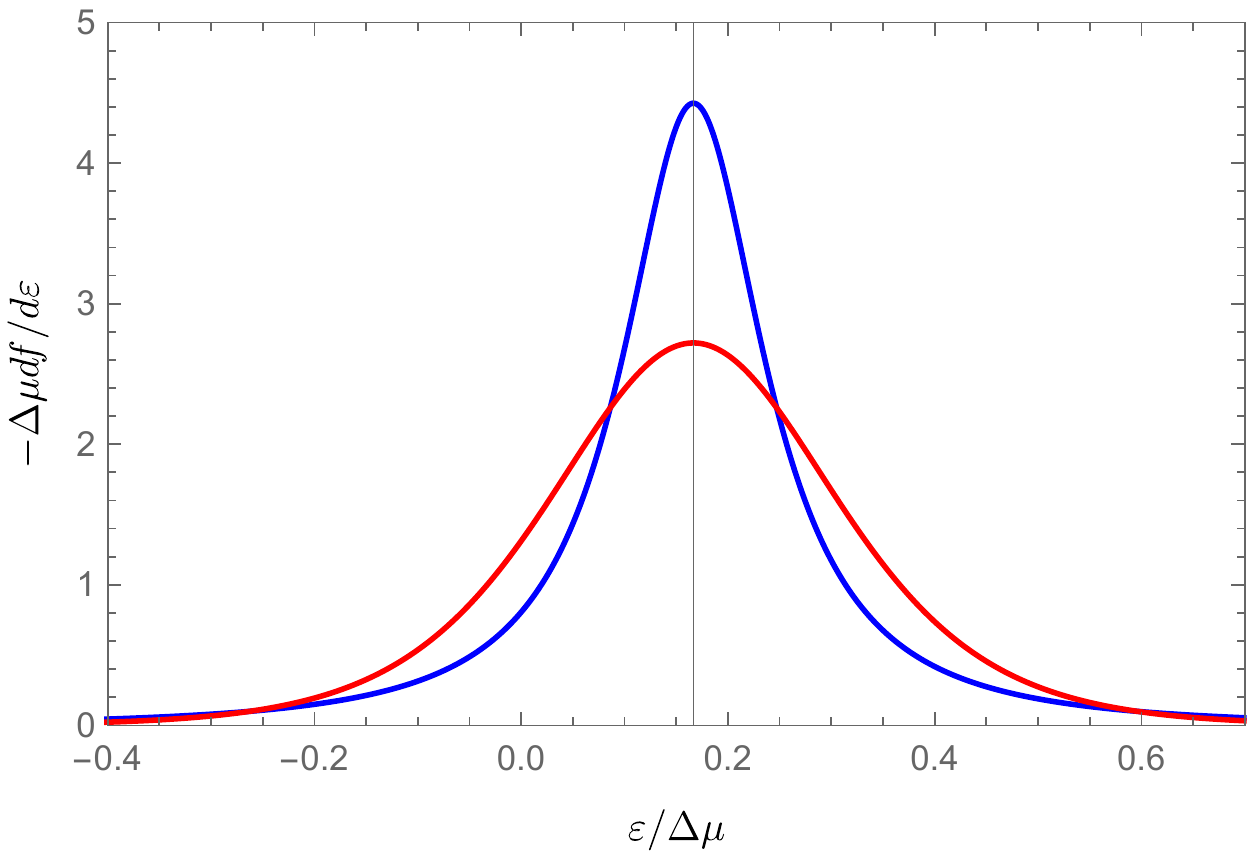}
\caption{\label{fig:two} Minus derivative of the distribution function in the outgoing channel (blue line), normalized to the voltage bias $\Delta\mu$, is plotted as a function of normalized energy for the load resistance $R=R_q/2$, and consequently, for $\lambda_{\rm in}=2\pi R/(R+R_q)=2\pi/3$. Broadening of the distribution function, caused by the Gaussian noise in the incoming channel, created by the QPC with the transparency $D=1/2$, is characterized by the effective temperature $T_{\rm out}$ in Eq.\ (\ref{efT}). It is found by comparing the energy flux in the channel, $J_{\rm out}=(R_q/2)\langle j_{\rm in}^2\rangle$, to the one for the equilibrium distribution (shown by the red line). Both distribution are shifted in energy by the value $\Delta \mu_{\rm out} = D \mathcal{T}_{\rm in} \Delta \mu=\Delta\mu/6$. Interestingly, the equilibrium distribution appears to be broader, which can be attributed to the fact that it saturates faster at high energies, than the none-equilibrium one.}
\end{figure}

In order to characterize the non-equilibrium distribution function in the outgoing channel, we compare it to the equilibrium one for the same effective temperature $T_{\rm out}$, which is determined as follows. The electronic energy flux of the equilibrium chiral channel reads
\begin{equation}
\label{fluxd}
J_{\rm out}= \frac{1}{R_q}\int\limits_{-\infty}^{\infty}d\varepsilon \varepsilon [f(\varepsilon)-\theta(\Delta \mu_{\rm out}-\varepsilon)]=\pi T^2_{\rm out}/12,
\end{equation}
which is nothing but the heat flux quantum. For the non-equilibrium channel, one can directly use the expression (\ref{result}) with the definition of the energy flux (\ref{fluxd}), compare the results, and thus find the effective temperature of the Gaussian noise. Alternatively, one can use the expression $J_{\rm out}=(R_q/2)\langle j_{\rm out}^2\rangle$ in terms of bosons (see, e.g., Ref.\ \cite{Slobodeniuk}), and compare it to the energy flux quantum (see the supplementary material). With both methods one obtains
\begin{equation}
\label{efT}
T^2_{\rm out}=\frac{3}{\pi^2}D(1-D)\left(\mathcal{T}_{\rm in}\Delta \mu\right)^2.
\end{equation}
One can now use this expression for the effective temperature $T_{\rm out}$ and the effective bias $\Delta \mu_{\rm out} = D \mathcal{T}_{\rm in} \Delta \mu $ of the channel to plot the equilibrium distribution function (see Fig.\ \ref{fig:two}, red line). Interestingly, the equilibrium distribution functions appears to be broader than the non-equilibrium one, which, we think, is compensated by its faster decay at large energies.

\textit{Effects of  heating.} So far, we have neglected the effect of  heating of the Ohmic contact assuming that its temperature vanishes, $T_{c}=0$, so that the Langevin sources $j_{c}$ and $j_{R}$ contain only ground-state fluctuations. This assumption greatly simplifies the operation of the quantum ammeter, however, it is justified only if coupling to phonons is sufficiently strong, so that they are able to evacuate extra heat. Now, we relax this requirement and assume full thermalization inside the Ohmic contact, so the temperature $T_{c}$ is to be found from the energy balance equations (for the recent closely related analysis, which accounts for coupling to phonons, see Ref.\ \cite{recent}). 

The incoming energy flux is given by $J_{\rm in}=(R_q/2)\langle j_{\rm in}^2\rangle$ (see, e.g., Ref.\ \cite{Slobodeniuk}), while the outgoing energy flux reads $J_{\rm out}=(R_q/2)\langle j_{\rm out}^2\rangle+(R/2)\langle {j'}^2_{\!\! {\rm out}}\rangle$. The total energy flux is conserved, $J_{\rm in}=J_{\rm out}$. By using Eqs.\ (\ref{Outgoing currents}) and (\ref{coefficients}),  we obtain $J_{\rm in}=(R_q/2)\langle j_{c}^2\rangle+(R^2/2R_q)\langle j_{R}^2\rangle$. By substituting the correlators (\ref{LS}) to this equation, integrating over $\omega$, and subtracting the ground-state contribution, we arrive at the following result:
\begin{equation}
\label{TC}
T_{c}^2=\frac{6R_q^2}{\pi(R_q+R)}\left[\langle j_{\rm in}^2\rangle-\langle j_{\rm in}^2\rangle_0\right],
\end{equation}
where $\langle j_{\rm in}^2\rangle_0$ is the ground-state part \cite{footnote}. Finally, one can use the temperature  $T_{c}$ from this equation in order to evaluate the correlators (\ref{LS}), and eventually, the distribution function (\ref{distrib}) with the help of Eqs.\ (\ref{Outgoing currents}-\ref{K}).

The difficulty that arises in the last step is that for the situations of interest, $R/ R_q\sim 1$, where the coupling to the charged mode is not small, ${\cal T}_{\rm in}\sim 1$, the neutral modes contribute an arbitrary fraction $q=\sqrt{1-{\cal T}^2_{\rm in}}$ of the equilibrium single boson channel at the temperature $T_{c}$ to the outgoing state, which is not small, either. Therefore, the FCS generator in Eq.\ (\ref{FCS}) is modified as follows:
\begin{equation}
\tilde {\cal G}(\lambda_{\rm in},t)= {\cal G}(\lambda_{\rm in},t)K_q(t, T_{c}),
\end{equation}
where $K_q(t, T_{c})=\langle e^{-iq\phi_{c}(t)}e^{iq\phi_{c}(0)}\rangle$ is the equilibrium correlator of the fractional quasiparticle with the charge $q\leq 1$ normalized to its ground state value, so that $K_q(t, 0)=1$. This correlator smears out the distribution function (\ref{distrib2}) introducing the effect that is difficult to account analytically. Therefore, we suggest to calibrate the quantum ammeter by measuring $-df/d\varepsilon$ for a noiseless biased incoming channel, $ j_{\rm in}=\langle j_{\rm in}\rangle$, and evaluating the Fourier transform to find the function $K_q(t, T_{c})$. Then, for the case of the non-zero noise, one can bias all the incoming channels in order to cancel the average current contribution in Eq.\ (\ref{TC}) to obtain $\langle \delta j_{\rm in}^2\rangle=\langle j_{\rm in}\rangle^2$. In this case, the function $K_q$ remains the same, and one can simply cancel it in the results of the measurements in order to obtain the FCS generator ${\cal G}(\lambda_{\rm in},t)$.

\textit{Discussion.}
First, we note, that in the limit $R\to\infty$ (floating Ohmic contact) $\lambda_{\rm in}=2\pi$, as follows from Eq.\ (\ref{coefficients1}). Heating is also negligible, according to Eq.\ (\ref{TC}). Therefore ${\cal G}$ is given by Eq.\ (\ref{FCS}) and is equal to the correlation function of electrons in the incoming channel. Thus, in this case the Ohmic contact not only conserves the phase of incoming electrons in the outgoing channel, as has been experimentally demonstrated in Ref.\ \cite{Pierre4}, but should also conserve the electron distribution function at all energies smaller than $E_c$. 

Second, throughout the paper we assumed that incoming and outgoing channels do not interact with other QH channels, if they are present. This assumption is justified, if the distance to the injection and detection point from the Ohmic contact is shorter than $v/\Delta\mu$, where $v$ is the velocity of the slowest neutral mode. In the opposite limit, one should take into account the charge fractionalization caused by the interaction between channels. At large distances one can neglect accumulated phases, which leads to the relatively simple modification of the scattering coefficients, Eq.\  (\ref{coefficients}). For instance, at the filling factor $\nu=2$ the strong Coulomb interaction splits the spectrum of the edge plasmons into the neutral and charged mode, and equally distributes each injected electron between two channels \cite{split}.

Finally, we have focused on the QH systems at integer filling factors, where the fermionic description is allowed. At fractional fillings the physics is more intricate. First of all, the spectrum of collective excitations at the edge is more complex, e.g., due to the emergence of upstream neutral mode \cite{upstream}. Second, the statistics of injected currents might be less simple at large transparencies. And what is more important, the physics of the Ohmic contact at general filling fractions is not yet well understood. However, at $\nu=1/m$, where $m$ is the odd integer number, edge of the QH system is known to support solely one charged mode. In this case, our model may still be applied by replacing $R_q\to mR_q$, and after some modification of the correlation functions \cite{IvanSingle}.

To summarize, we have investigated an Ohmic contact, a small piece of metal perfectly coupled to chiral QH edge states, and demonstrated that at energies small compared to its charging energy $E_c$, it fractionalizes electron states, thus only a fraction ${\cal T}_{\rm in}$ of an incoming electron is transmitted to one of the outgoing channels. We have shown that this system can serve as a ``quantum ammeter'' device: By monitoring the resonant tunneling current through a QD attached to the outgoing channel one can measure the generator of the FCS of the  incoming current, with $\lambda_{\rm in}=2\pi {\cal T}_{\rm in}$ playing the role of the counting variable. After presenting several applications of this device, we have  discussed how to calibrate the quantum ammeter to account for the effects of heating. 

We are grateful to F. Pierre for fruitful discussion. E. I. acknowledges financial support from the Fonds National de la Recherche Luxembourg under the grants AT-
TRACT 7556175 and INTER 11223315. E. S. acknowledges financial support from the Swiss National Science Foundation.


\begin{thebibliography}{99}

\bibitem{Klitzing}
K. v. Klitzing, G. Dorda, and M. Pepper, Phys. Rev. Lett. {\bf 45}, 494 (1980).

\bibitem{Tsui}
D. C. Tsui, H. L. Stormer, and A. C. Gossard, Phys. Rev. Lett. {\bf 48}, 1559 (1982).

\bibitem{Ezawa}
Z. F. Ezawa, {\em Quantum Hall Effects: Recent Theoretical
and Experimental Developments,} (3rd ed., World Scientific,
2013).

\bibitem{Laughlin}
R. B. Laughlin, Phys. Rev. Lett. {\bf 50}, 1395 (1983).

\bibitem{Reznikov}
R. de-Picciotto, M. Reznikov, M. Heiblum, V. Umansky, G. Bunin, and D. Mahalu, Nature {\bf 389}, 162 (1997).

\bibitem{Saminadayar}
L. Saminadayar, D. C. Glattli, Y. Jin, and B. Etienne, Phys. Rev. Lett. {\bf 79}, 2526 (1997).

\bibitem{Ji}
Y. Ji, Y. Chung, D. Sprinzak, M. Heiblum, D. Mahalu, and H. Shtrikman, Nature {\bf 422}, 415 (2003);
E. Bieri, M. Weiss, O. Goktas, M. Hauser, S. Csonka, S. Oberholzer, C. Schonenberger, Phys. Rev. B {\bf 79}, 245324 (2009); L. V. Litvin, H.-P. Tranitz, W. Wegscheider, and C. Strunk, Phys. Rev. B {\bf 75}, 033315 (2007); I. Neder, F. Marquardt, M. Heiblum, D. Mahalu, and V. Umansky, Nat Phys {\bf 3}, 534 (2007); P. Roulleau, F. Portier, D. C. Glattli, P. Roche, A. Cavanna, G. Faini, U. Gennser, and D. Mailly, Phys. Rev. B {\bf 76}, 161309 (2007); E. Bieri, M. Weiss, O. Goktas, M. Hauser, S. Csonka, S. Oberholzer, C. Schonenberger, Phys. Rev. B {\bf 79}, 245324 (2009); H. Duprez, E. Sivre, A. Anthore, A. Aassime, A. Cavanna, A. Ouerghi, U. Gennser, and F. Pierre, Phys. Rev. X {\bf 9}, 021030 (2019).

\bibitem{PPPFujisawa}
C. Altimiras, H. le Sueur, U. Gennser, A. Cavanna, D. Mailly, and F. Pierre, Nature Physics {\bf 6}, 34 (2010); H. le Sueur, C. Altimiras, U. Gennser, A. Cavanna, D. Mailly, and F. Pierre, Phys. Rev. Lett. { \bf 105}, 056803 (2010);
C. Altimiras, H. le Sueur, U. Gennser, A. Cavanna, D. Mailly, and F. Pierre,
Phys. Rev. Lett. { \bf 105}, 226804 (2010); K. Itoh, R. Nakazawa, T. Ota, M. Hashisaka, K. Muraki, and T. Fujisawa, Phys. Rev. Lett. {\bf 120}, 197701 (2018).


\bibitem{Degiovanni} For a review, see
C. Grenier, R. Herv\'e, Gwendal F\'eve, and P. Degiovanni, Mod. Phys. Lett. B {\bf 25}, 1053 (2011).


\bibitem{Buttiker}
M. B\"{u}ttiker, IBM J. Res. Develop. {\bf 32}, 63 (1988).

\bibitem{Pierre1}
S. Jezouin, Z. Iftikhar, A. Anthore, F. D. Parmentier, U. Gennser, A. Cavanna, A. Ouerghi, I. P. Levkivskyi, E. Idrisov, E. V. Sukhorukov, L. I. Glazman, and F. Pierre, Nature {\bf 536}, 58 (2016).


\bibitem{Idrisov1}
E. G. Idrisov, I. P. Levkivskyi, and E. V. Sukhorukov, Phys. Rev. B {\bf 96}, 155408 (2017).

\bibitem{Slobodeniuk}
A. O. Slobodeniuk, I. P. Levkivskyi, and E. V. Sukhorukov, Phys. Rev. B {\bf 88}, 165307 (2013).

\bibitem{Pierre2}
E. Sivre, A. Anthore, F. D. Parmentier, A. Cavanna, U. Gennser, A. Ouerghi, Y. Jin, and F. Pierre, Nature Physics {\bf 14}, 145 (2018).



\bibitem{Idrisov2}
E. G. Idrisov, I. P. Levkivskyi, and E. V. Sukhorukov, Phys. Rev. Lett., {\bf 121}, 026802 (2018).

\bibitem{Clerk}
A. A. Clerk, P. W. Brouwer, and V. Ambegaokar, Phys. Rev. Lett. {\bf 87}, 186801 (2001).

\bibitem{Pierre4}
H. Duprez, E. Sivre, A. Anthore, A. Aassime, A. Cavanna, U. Gennser, F. Pierre, arXiv:1902.07569. 

\bibitem{Matveev}
A. Furusaki and K. A. Matveev, Phys. Rev. B {\bf 52}, 16676 (1995).

\bibitem{Pierre3}
Z. Iftikhar, A. Anthore, A. K. Mitchell, F. D. Parmentier, U. Gennser, A. Ouerghi, A. Cavanna, C. Mora, P. Simon, and F. Pierre, Science (2018): {\bf eaan5592}, DOI: 10.1126/science.aan5592.

\bibitem{Heiblum1}
M. Banerjee, M. Heiblum, A. Rosenblatt, Y. Oreg, D. E. Feldman, A. Stern, and V. Umansky, Nature {\bf 545}, 75 (2017).

\bibitem{Heiblum2}
M. Banerjee, M. Heiblum, V. Umansky, D. E. Feldman, Y. Oreg, and A. Stern, 
Nature {\bf 559}, 210 (2018).

\bibitem{Levitov}
L. S. Levitov, H. Lee, and G. B. Lesovik, J. Math. Phys {\bf 37}, 4845 (1996).

\bibitem{footnote0}
In the long time limit  $Q_{\rm in}(0) $ and $Q_{\rm in}(t) $ commute, and ${\cal G}$ becomes the classical moment generating function of the injected charge $\Delta Q_{\rm in}(t)=Q_{\rm in}(t)-Q_{\rm in}(0)$, i.e., $\partial^m_{i\lambda}{\cal G}(i\lambda)|_{\lambda=0}=\langle (\Delta Q_{\rm in})^m\rangle$.


\bibitem{Wen}
X. G. Wen, Phys. Rev. B {\bf 41}, 12838 (1990).

\bibitem{Giamarchi}
T. Giamarchi, {\em Quantum Physics in One Dimension}
(Claverdon Press Oxford, 2004).

\bibitem{Sukhorukov1}
E. V. Sukhorukov, Physica E {\bf 77}, 191 (2016).

\bibitem{Chernii}
I. Chernii, I. P. Levkivskyi, E. V. Sukhorukov,
 Phys. Rev. B {\bf 90}, 245123 (2014).

\bibitem{Borin}
A. Borin, E. Sukhorukov, Phys. Rev. B {\bf 99}, 085430 (2019).

\bibitem{deJong} M. J. M. de Jong, Phys. Rev. B {\bf 54}, 8144 (1996).

\bibitem{Levkivskyi}
I. P. Levkivskyi, and E. V. Sukhorukov,
Phys. Rev. Lett. {\bf 103}, 036801 (2009).

\bibitem{recent}
E. Sivre, H. Duprez, A. Anthore, A. Aassime, F.D.
Parmentier, A. Cavanna, A. Ouerghi, U. Gennser, and F. Pierre, unpublished.

\bibitem{footnote}
Note, that according to Eq.\ (\ref{TC}) $T_c^2$ can be used as a measure  of instant (coincedent in time) fluctuations of the current.

\bibitem{split}
I. P. Levkivskyi, and E. V. Sukhorukov, Phys. Rev. B {\bf 78}, 045322 (2008).

\bibitem{upstream} A. H. MacDonald, Phys. Rev. Lett. {\bf 64}, 220 (1990); M. D. Johnson and A. H. MacDonald, Phys. Rev. Lett. {\bf 67}, 2060 (1991); C. L. Kane and M. P. A. Fisher, Phys. Rev. B {\bf 55}, 15832 (1997).

\bibitem{IvanSingle}
Ivan P. Levkivskyi, Phys. Rev. B {\bf 93}, 165427 (2016)

\end{thebibliography}
\end{document}


\title{ Quantum ammeter. Supplementary material.}

\author{Edvin G. Idrisov}
\affiliation{Physics and Materials Science Research Unit, University of Luxembourg, L-1511 Luxembourg, Luxembourg}

\author{Ivan P. Levkivskyi}
\affiliation{Dropbox Ireland, One Park Place, Hatch Street Upper, Dublin, Ireland}

\author{Eugene V. Sukhorukov}
\affiliation{D\'epartement de Physique Th\'eorique, Universit\'e de Gen\`eve, CH-1211 Gen\`eve 4, Switzerland}
\date{\today}




\maketitle


\section{Langevin equations and electron distribution function}

Here, we follow the Refs.\ \cite{Slobodeniuk,Idrisov1,Idrisov2} and write the equations of motion for the currents and the charge $Q$ in the Ohmic contact in the  form (see Fig.\ 1 of the main text):
\begin{equation}
 \dot{Q}(t)=j_{\rm in}(t)-j_{\rm out}(t)-j'_{\rm out}(t), \quad
 j_{\rm out}(t)=\frac{Q(t)}{R_q C}+j_c(t), \quad
j^{\prime}_{\rm out}(t)= \frac{Q(t)}{R C}+j_R(t),
\end{equation}
where the first equation expresses the conservation of charge, while the second and the third one are the Langevin equations, which have the following physical meaning: On the right hand side the contributions $Q(t)/R_q C$ and $Q(t)/R C$ are the currents induced by the time-dependent potential $Q(t)/C$ of the Ohmic contact, while the terms  $j_c(t)$ and $j_R(t)$ are the Langevin current sources. After straightforward calculations, we present the outgoing currents in the following form
\begin{equation}
j_{\rm out}(\omega)= \sum_{\alpha={\rm in},c,R} \mathcal{T}_\alpha(\omega)j_\alpha(\omega),\quad j'_{\rm out}(\omega)= \sum_{\alpha={\rm in},c,R} \mathcal{T}'_\alpha(\omega)j_\alpha(\omega)
\end{equation} 
where $\mathcal{T}_R(\omega)=-\mathcal{T}_{\rm in}(\omega)=\mathcal{T}_c(\omega)-1=1/[i\omega R_q C-R_q/R -1]$, and $\mathcal{T}'_c(\omega)=-\mathcal{T}'_{\rm in}(\omega)=\mathcal{T}'_R(\omega)-1=(R_q/R)\mathcal{T}_R(\omega)$. At frequencies small compared to the charging energy, $\omega \ll E_c\sim 1/R_qC$, the outgoing currents are given by Eq.\ (3) of the main text.

Next, we use the Eq.\  (5) of the main text to obtain the following expression for the electron correlation function at zero temperature: 
\begin{equation}
 K(t) \propto (it+0)^{-2R_q {\cal T}^2_c G_c-2R_q {\cal T}^2_R G_R} \langle e^{i\lambda_{\rm in} Q_{\rm in}(t)}e^{-\lambda_{\rm in}Q_{\rm in}(0)}\rangle.
\end{equation}
We note, that the coefficients ${\cal T}_\alpha$ satisfy the summation rule:  $\sum_\alpha |{\cal T}_\alpha|^2G_\alpha=1/2R_q $, where $G_{\rm in}=G_c=1/2R_q$, $G_R=1/R$. Taking this relation into account and using Eq.\ (2), we straightforwardly get Eqs.\ (7) of main text, namely 
\begin{equation}
-\frac{df(\varepsilon)}{d\varepsilon} = \int \frac{dt}{2\pi} e^{-i\varepsilon t}\mathcal{G}(\lambda_{\rm in},t), \quad \text{where} \quad \mathcal{G}(\lambda_{\rm in},t ) \propto (it+0)^{{\cal T}^2_{\rm in}} \langle e^{i\lambda_{\rm in} Q_{\rm in}(t)}e^{-\lambda_{\rm in}Q_{\rm in}(0)}\rangle.
\end{equation}

\section{Full counting statistics of sequential tunneling}

\begin{figure}
\includegraphics[scale=.6]{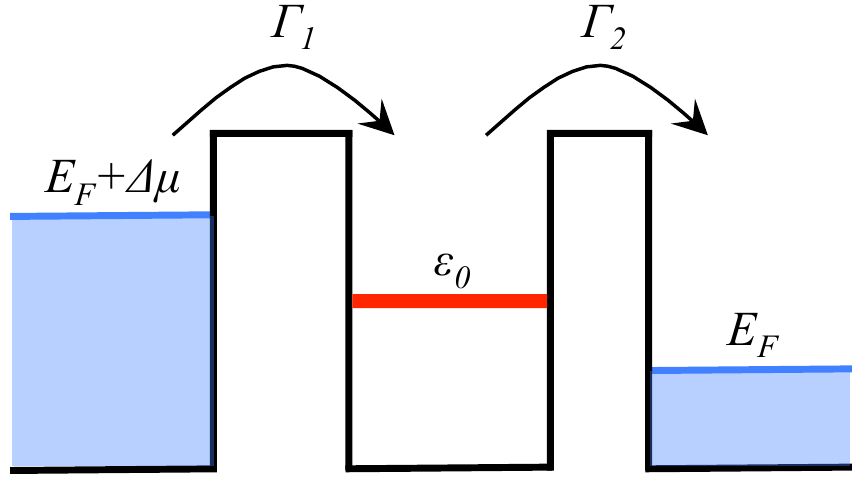}
\caption{\label{fig:one:a}  Sequential tunneling via the resonant level $\varepsilon_0$ in a double-barrier structure, e.g., in a QD, is schematically shown. At strong biases, $E_F+\Delta\mu>\varepsilon_0>E_F$ and $\Delta\mu\gg\Gamma_{1,2}$, the transport of electrons arises due to tunneling from the left to the right reservoir through two barriers with the rates $\Gamma_1$ and $\Gamma_2$.}
\end{figure}

According to Ref.\ \cite{deJong}, the FCS generator of sequential electron tunneling through a single level in a QD can be  obtained by solving  the generalized master equation. Here, we briefly derive it by considering the evolution of the joint probability distribution function $P_l(n,t)$, where $l=1,2$ enumerates states of the empty and occupied level, and $n$ is the number of electrons in the right reservoir (see Fig.\ \ref{fig:one:a}).  At zero temperature, the probability distribution function satisfies the infinite set of master equations: 
\begin{equation}
\label{master1}
\partial_t P_1(n,t)=-\Gamma_1 P_1(n,t)+\Gamma_2 P_2(n-1,t),\quad \partial_t P_2(n,t)=-\Gamma_2 P_2(n,t)+\Gamma_1 P_1(n,t).
\end{equation}
Then, the FCS generator can formally be written as:
\begin{equation}
\label{generators}
{\cal G}(\lambda_{\rm in},t)=\sum_{l=1,2}g_l(\lambda_{\rm in},t),\quad \text{where}\quad g_l(\lambda_{\rm in},t)=
\sum_{n,l',n'}P_l(n,t)P^{(0)}_{l'}(n')e^{i\lambda_{\rm in}(n-n')},
\end{equation}
with the initial condition $P_l(n,0)=\delta_{l,l'}\delta_{n,n'}$, while $P^{(0)}_l(n)$ is the stationary distribution of master equations (\ref{master1}). 

We have intentionally introduced the moment generator $g_l(\lambda_{\rm in},t)$ in the mixed representation, because it satisfies simple generalized master equations
\begin{equation}
\label{master2}
\partial_t g_1(\lambda_{\rm in},t)=-\Gamma_1 g_1(\lambda_{\rm in},t)+\Gamma_2e^{i\lambda_{\rm in}} g_2(\lambda_{\rm in},t),\quad \partial_t g_2(\lambda_{\rm in},t)=-\Gamma_2 g_2(\lambda_{\rm in},t)+\Gamma_1 g_1(\lambda_{\rm in},t),
\end{equation}
as follows from its definition and from the master equations  (\ref{master1}). The formal solution of these equations reads:
\begin{equation}
g_l(\lambda_{\rm in},t)=\sum_{m=1,2}c_mv_{ml}e^{i\Lambda_m t}, \quad t>0,
\end{equation}
where $v_{ml}$ and $\Lambda_m$ are the eigenvectors and eigenvalues of the matrix of the coefficients in Eq.\ (\ref{master2}), respectively, and $c_k$ are arbitrary constants. These coefficients can be found from the initial condition for $g_l(\lambda_{\rm in},t)$, which follows from Eq.\ (\ref{generators}), that it is equal to the stationary occupations of the QD level: $g_1=\Gamma_2/(\Gamma_1+\Gamma_2)$ and   
$g_2=\Gamma_1/(\Gamma_1+\Gamma_2)$. 

Taking the final trace, ${\cal G}(\lambda_{\rm in},t)=\sum_{l=1,2}g_l(\lambda_{\rm in},t)$, after straightforward calculations we arrive at the following result 
\begin{equation}
\mathcal{G}(\lambda_{\rm in},t)=\sum_{m=1,2}\mathcal{G}_m(\lambda_{\rm in})e^{\Lambda_m(\lambda_{\rm in})t},\quad t>0,
\end{equation}
where the  ``weights'' and eigenvalues are given by $\mathcal{G}_{1,2}(\lambda_{\rm in})=[1 \pm F(\lambda_{\rm in})]/2$ and $\Lambda_{1,2}(\lambda_{\rm in})=-[\Gamma_1+\Gamma_2 \mp \beta(\lambda_{\rm in}) ]/2$, respectively, with
\begin{equation}
\begin{split}
& F(\lambda_{\rm in})=\frac{\beta(\lambda_{\rm in})}{\Gamma_1+\Gamma_2}-\frac{2\Gamma_1 \Gamma_2 (e^{i\lambda_{\rm in}}-1)}{\beta(\lambda_{\rm in})[\Gamma_1+\Gamma_2]}, \\
& \beta(\lambda_{\rm in})=\sqrt{(\Gamma_1+\Gamma_2)^2+4\Gamma_1 \Gamma_2 (e^{i\lambda_{\rm in}}-1)}. \\
\end{split}
\end{equation}

Finally, using Eqs.\ (7) of the main text the electron distribution function can be presented in the following form 
\begin{equation}
-\frac{d f(\varepsilon)}{d\varepsilon}=\text{Re}\left[ \int^{\infty}_0 \frac{dt}{\pi}e^{-i\varepsilon t} \mathcal{G}(\lambda_{\rm in},t)\right]=\sum_{m=1,2} \text{Re}\left[ \frac{1}{\pi}\frac{\mathcal{G}_m}{\gamma_m+i(\varepsilon-\varepsilon_m)}\right],
\end{equation} 
where  $\varepsilon_m = \text{Im} [\Lambda_m (\lambda_{\rm in})]$ and $\gamma_m = -\text{Re}[\Lambda_m (\lambda_{\rm in})]>0$. For the case of the symmetric system, $\Gamma_1=\Gamma_2=\Gamma$, these expressions take the simple form: $\Lambda_{1,2}=-\Gamma \pm \Gamma e^{i\lambda_{\rm in}/2}$ and $\mathcal{G}_{1,2}=[1 \pm \cos(\lambda_{\rm in}/2)]/2$. 

\section{Broadening  of the electron distribution function by a  Gaussian noise}
Typically, the effective noise temperature is defined by comparing the zero-frequency noise power to its equilibrium value given by the fluctuation-dissipation relation. However, in order to compare equilibrium and non-equilibrium distribution functions, it is more appropriate to define the effective noise temperature via the energy fluxes of two distributions (for the details, see Ref.\ \cite{Slobodeniuk}), which characterize their broadening. The total energy flux in the outgoing channel can be written as
\begin{equation}
J_{\rm out}=(R_q/2)[\langle j^2_{\rm out}\rangle-\langle j^2_{\rm out}\rangle_{0}]=(1/2)\int d\omega [S_{\rm out}(\omega)-S_{0}(\omega)],
\end{equation}
where
\begin{equation}
S_{\rm out}(\omega)=\int dt e^{i\omega t}\langle j_{\rm out}(t)j_{\rm out}(0)\rangle
\end{equation}
is the noise power of the outgoing current, and $S_0(\omega)$ is its ground-state value. For the equilibrium chiral channel with the temperature $T_{\rm out}$ the fluctuation-dissipation relation reads:   
\begin{equation}
S_{\rm out}(\omega)=\frac{\omega/R_q}{1-e^{-\omega/T_{\rm out}}},\quad \text{ and}\quad  S_0(\omega)=\omega \theta(\omega)/R_q,
\end{equation}
thus the equilibrium channel carries the energy flux $J_{\rm out}=\pi T^2_{\rm out}/12$, which is nothing but the flux quantum.

In the case of a Gaussian non-equilibrium noise, one can use Eq.\ (3) of the main text and the property $\sum_{\alpha} \mathcal{T}^2_{\alpha} G_{\alpha}=1/2R_q$, where $G_{\rm in}=G_c=1/2R_q$ and $G_R=1/R$, to write
\begin{equation}
\label{eq1}
J_{\rm out}=\frac{1}{2}\sum_{\alpha={\rm in}, c,R} \int d\omega\left[  \mathcal{T}^2_{\alpha}  S_{\alpha}(\omega)-S_0(\omega) \right]=
(\mathcal{T}^2_{\rm in}/2) \int d\omega\left[S_{\rm in}(\omega)-S_0(\omega)\right],
\end{equation}
where $S_{\rm in}(\omega)$ is the noise power in the incoming channel.  By applying the scattering theory \cite{BB}, one obtains for the voltage biased QPC at zero temperature:
\begin{equation}
\label{eq2}
S_{\rm in}(\omega)=S_0(\omega)+D(1-D)\times\left[\sum_{\sigma=\pm}S_{0}(\omega+\sigma \Delta \mu)-2S_0(\omega)\right].
\end{equation}
Substituting this expression into Eq.\ (\ref{eq1}), one easily obtains
\begin{equation}
\label{Heat flux}
J_{\rm out}=\frac{1}{4\pi}D(1-D)\left(\mathcal{T}_{\rm in}\Delta \mu\right)^2
\end{equation} 
Interestingly, for $\mathcal{T}_{\rm in}=1$ this expression agrees with the electron energy flux evaluated right after the QPC. Thus, the suppression of high-order cumulants conserves the energy flux.

\begin{figure}
\includegraphics[scale=.7]{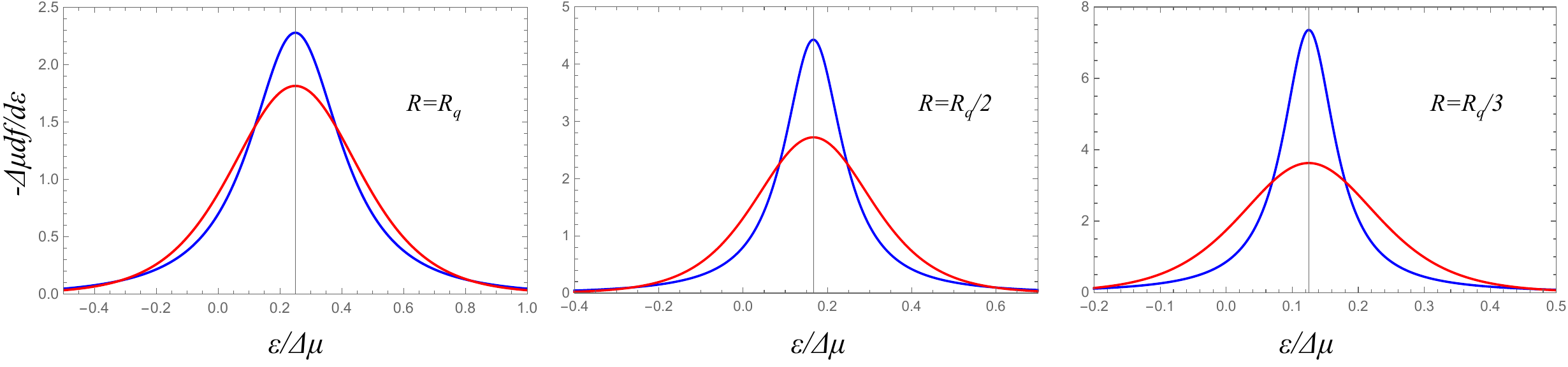}
\caption{\label{fig:two:a}   Minus derivative of the distribution function (blue lines), normalized to the voltage bias $\Delta\mu$, is plotted as a function of normalized energy for different values of the load resistance $R$ [consequently, for different coupling constants $\lambda_{\rm in}=2\pi R/(R+R_q)$]. Broadening of the distribution function, caused by the Gaussian noise of the QPC with the transparency $D=1/2$, is characterized by the effective temperature in Eq.\ (\ref{efT}), evaluated by comparing the energy flux in the channel to the one for the equilibrium distribution (red lines).}
\end{figure}

Finally, we compare the result  (\ref{Heat flux}) to the flux quantum $J_{\rm out}=\pi T_{\rm out}^2/12$ to defind the effective temperature of the channel
\begin{equation}
\label{efT}
T^2_{\rm out}=\frac{3}{\pi^2}D(1-D)\left(\mathcal{T}_{\rm in}\Delta \mu\right)^2.
\end{equation}
Note, that only the fraction $D\mathcal{T}_{\rm in} $ of electrons from the reservoir reaches the outgoing channel. Thus, the effective bias in this channel reads:
\begin{equation}
\label{bias}
\Delta \mu_{\rm out} = D\mathcal{T}_{\rm in} \Delta \mu .
\end{equation}

In the next step, we use Eqs.\ (7) of the main text, the Gaussian character of the noise (expanding the correlator in Eqs.\ (7) to second order in $\lambda_{\rm in}$, averaging, and re-exponentiating the result), and Eq.\ (\ref{eq2}), to obtain
\begin{equation}
\label{eq3}
-\frac{\partial f}{\partial \varepsilon}= \int \frac{dt}{2\pi}e^{-i\varepsilon t} \mathcal{G}_{\rm in}(\lambda_{\rm in},t), \quad \log \mathcal{G} = it \Delta \mu D \mathcal{T}_{\rm in}-2D(1-D)\mathcal{T}^2_{\rm in}F(t\Delta \mu),
\end{equation}
where $F(t\Delta\mu)=\int_0^1dx(1-x)x^{-2}[1-\cos(t\Delta\mu x)]$ is the dimensionless function, and we set $\lambda_{\rm in}=2\pi \mathcal{T}_{\rm in}$. We are now in the position to compare the equilibrium and non-equilibrium distribution function. For doing so, we evaluate the integral in Eq.\ (\ref{eq3}) numerically and present the results in Fig.\ \ref{fig:two:a} together with the equilibrium distributions with the temperature $T_{\rm out}$ and the bias $\Delta\mu_{\rm out}$ from Eqs.\ (\ref{efT}) and (\ref{bias}) for different values of $\lambda_{\rm in}$. The apparent difference in broadening of equilibrium and non-equilibrium distribution functions (with the same effective temperatures) can, most likely, be attributed to the fact, that the non-equilibrium distribution decays slower at high energies. This effect is more pronounced for smaller values of $\lambda_{\rm in}$, because at $\lambda_{\rm in}\ll 1$ the function $-df/d\varepsilon$ acquires power-law tails at $|\varepsilon|\gg\Delta\mu$.